\documentclass{PoS}

\usepackage{txfonts}

\let\OLDthebibliography\thebibliography
\renewcommand\thebibliography[1]{
  \OLDthebibliography{#1}
  \setlength{\parskip}{0pt}
  \setlength{\itemsep}{0pt}
}

\newcommand{\ie}{{\it i.e.}}
\newcommand{\eg}{{\it e.g.}}

\def\AFTER {\mbox{AFTER@LHC}\xspace}	
\usepackage{xspace}
\usepackage{comment}

\usepackage{subfigure}

\usepackage{lineno}

\newcommand{\cf}[1]{{Fig.~\ref{#1}}}

\title{Single-Transverse-Spin-Asymmetry studies with a fixed-target experiment using the LHC beams (\AFTER)}

\ShortTitle{STSA studies at AFTER@LHC}

\author{\small {J.P.~Lansberg}$^1$\thanks{Speaker. Email:  Jean-Philippe.Lansberg@in2p3.fr}, 
M.~Anselmino$^2$, 
R.~Arnaldi$^2$, 
S.J.~Brodsky$^3$, 
V.~Chambert$^1$, 
C.~Da Silva$^4$,
J.P.~Didelez$^1$, 
M.G~Echevarria$^5$, 
E.G.~Ferreiro$^6$, 
F.~Fleuret$^7$,  
Y.~Gao$^{8}$, 
B.~Genolini$^1$, 
C.~Hadjidakis$^1$, 
I.~H\v{r}ivn\'{a}\v{c}ov\'{a}$^{1}$,
D. Kikola$^{9}$,
A.~Klein$^{4}$,
A. Kurepin$^{10}$,
A.~Kusina$^{11}$, 
C.~Lorc\'e$^{12}$,
F.~Lyonnet$^{13}$,
L.~Massacrier$^{1}$, 
A. Nass$^{14}$, 
C.~Pisano$^{15}$,
P.~Robbe$^{16}$,
I.~Schienbein$^{11}$,
M.~Schlegel$^{17}$,
E.~Scomparin$^2$,
J.~Seixas$^{18}$,
H.S.~Shao$^{19}$,
A.~Signori$^{20}$,
E.~Steffens$^{21}$,
N.~Topilskaya$^{10}$,
B.~Trzeciak$^{22}$
U.I.~Uggerh\o j$^{23}$, 
A.~Uras$^{24}$,
R.~Ulrich$^{25}$, 
and Z.~Yang$^{8}$.
\\ \footnotesize
$^1$ IPNO, Univ. Paris-Sud, CNRS/IN2P3, Universit\'e Paris-Saclay,  Orsay, France\\
$^2$ Dip. di Fisica and INFN Sez. Torino, Via P. Giuria 1, Torino, Italy \\
$^3$ SLAC National\,Accelerator\,Laboratory, Stanford University, Menlo Park, USA\\
$^4$ LANL, P-25, Los Alamos National Laboratory, Los Alamos, NM 87545, USA \\
$^5$ ECM, Universitat de Barcelona,  Barcelona, Spain\\
$^6$ Dept. de F{\'\i}sica de Part{\'\i}culas, USC, Santiago de Compostella, Spain\\
$^7$ LLR, \'Ecole Polytechnique, CNRS/IN2P3,   Palaiseau, France\\
$^{8}$ CHEP, Department of Engineering Physics, Tsinghua University, Beijing, China\\
$^{9}$ Faculty of Physics, Warsaw University of Technology,  Warsaw, Poland \\
$^{10}$ Institute for Nuclear Research, Russian Academy of Sciences, Moscow, Russia\\
$^{11}$ LPSC, Universit\'e Grenoble-Alpes, CNRS/IN2P3, 38026 Grenoble, France\\
$^{12}$ CPhT, Ecole Polytechnique, CNRS, Universit\'e Paris-Saclay,  Palaiseau, France\\
$^{13}$ Southern Methodist University, Dallas, TX 75275, USA\\
$^{14}$ Institut f\"ur Kernphysik, Forschungszentrum J\"ulich, J\"ulich, Germany \\
$^{15}$ Dipartimento di Fisica, Universita degli Studi di Pavia, Pavia, Italy\\
$^{16}$ LAL, Univ. Paris-Sud, CNRS/IN2P3, Universit\'e Paris-Saclay,  Orsay, France\\
$^{17}$ Institute for Theoretical Physics, T\"ubingen U.,  T\"ubingen, Germany \\
$^{18}$ LIP and IST, Lisbon, Portugal \\
$^{19}$ Theoretical Physics Department, CERN, CH-1211 Geneva 23, Switzerland \\
$^{20}$ Nikhef and Dept. of Physics and Astronomy, VU  Amsterdam,  Amsterdam, The Netherlands\\
$^{21}$ Physics Institute, Friedrich-Alexander University Erlangen-N\"urnberg, Erlangen, Germany\\
$^{22}$ Institute for Subatomic Physics, Utrecht University, Utrecht, The Netherlands\\
$^{23}$ Department of Physics and Astronomy, University of Aarhus, Denmark\\
$^{24}$ IPNL, Universit\'e Claude Bernard Lyon-I, CNRS/IN2P3, Villeurbanne, France\\
$^{25}$ Institut f\"ur Kernphysik, Karlsruhe Institute of Technology (KIT), Karlsruhe, Germany
       }

\abstract{We discuss the potential of AFTER@LHC  to measure single-transverse-spin
 asymmetries in open-charm and bottomonium production. With a HERMES-like hydrogen 
polarised target, such measurements over a year can reach precisions close to 
the per cent level. This is particularly remarkable since these analyses 
can probably not be carried out
anywhere else.}

\FullConference{XXIV International Workshop on Deep-Inelastic Scattering and Related Subjects\\
		11-15 April, 2016, 
		DESY Hamburg, Germany}

\begin{document}

\section{Introduction}

Continuous efforts have been made in the last decades to advance our knowledge 
of the internal structure of the nucleon and, in particular, that of the 
constituent dynamics, the quarks and the gluons. However, it remains largely 
unknown with a limited understanding of the proton and neutron spin structure, 
namely how they bind into a spin-$\frac{1}{2}$ object. 
There are two types of quark and gluon contributions to the nucleon spin: 
their spin and their Orbital Angular Momentum (OAM). 
For a $+\frac{1}{2}$ helicity nucleon, one has
$\frac{1}{2} = \frac{1}{2}\Delta \Sigma + \Delta G + {\cal L}_{g} + {\cal L}_{q}$
where $\frac{1}{2}\Delta\Sigma$ refers to the combined spin contribution
 of quarks and antiquarks, $\Delta G$ the gluon spin, and ${\cal L}_{q,g}$ 
the quark and gluon OAM contributions (see \eg\ \cite{Leader:2013jra,Wakamatsu:2014zza}). 

This equation applies whatever the energy scale we look at the nucleons
and this strongly motivates the scale-evolution study of the individual contributions.

The spin crisis
of the 80's has evolved into a puzzle, \ie\ to determine
how large the  quark and gluon contributions to the nucleon spin are, 
to disentangle them and eventually to explain them from first principles in QCD.
Recent experimental analyses pointed at $\Delta \Sigma$ as low as 0.25 ~\cite{deFlorian:2008mr} 
and that $\Delta G$ could\footnote{The latter has only
been probed for $x > 0.05$~\cite{Adamczyk:2014ozi,deFlorian:2014yva}} reach 0.2.
There is still ample room --rather, need-- for ${\cal L}_{q}$ and ${\cal L}_{g}$, 
which have not yet been measured. This highlights how important 
studies of the transverse motion of quarks and gluons inside the proton are.
Indirect\footnote{To measure the parton OAM, observables which are sensitive to the parton position
and momentum are in principle required. A well known example of such objects are 
Generalised Parton Distributions (GPDs) accessible in exclusive processes.} information on the orbital motion of the partons bound inside hadrons 
can be accessed via Single (Transverse) Spin Asymmetries (S(T)SA) 
in different hard-scattering processes, in particular with a transversely polarised hadron 
(see~\cite{D'Alesio:2007jt,Barone:2010zz} for recent reviews) to probe  
the Sivers effect~\cite{Sivers:1989cc,Sivers:1990fh} \footnote{It is connected to left-right asymmetries 
in the parton distributions with respect to the plane formed by the proton 
momentum and spin directions}. This effect is naturally connected~\cite{Burkardt:2002ks,Burkardt:2003uw} to the 
transverse motion of the partons inside the polarised nucleons (see also \cite{Liu:2015xha}). 

As of today, the Sivers effect can be approached via two dual formalisms~\cite{Koike:2007dg}. 
One extends the collinear parton model of Bjorken with quark-gluon or gluon-gluon 
correlation functions~\cite{Efremov:1981sh,Efremov:1984ip,Qiu:1991pp}. 
It is called the Collinear Twist-3 (CT3) formalism. Another, referred to as the 
Transverse-Momentum Dependent (TMD) factorisation (see~\eg~\cite{Collins:2011zzd,GarciaEchevarria:2011rb,Angeles-Martinez:2015sea}), uses a complete tri-dimensional mapping 
of the parton momentum. It can also deal with all the possible spin-spin and 
spin-orbit correlations between the hadron and the partons.
Both approaches encode the re-scatterings\footnote{They are accounted by gauge 
links in the definition of the TMD PDFs and explicitly considered in the 
hard-scattering coefficient for CT3.} of the quarks and gluons with the hadron 
remnants~\cite{Brodsky:2002rv,Collins:2002kn,Brodsky:2002cx} which 
are believed to generate the STSAs, also referred to $A_N$. 
Both formalisms have their preferred range of applicability.
As for the CT3 observables, the \AFTER physics case adds up 
heavy-flavour and bottomonium production, whose STSAs are essentially unknown, 
to the usual list of $A_N$ studies which includes single hadron~\cite{Qiu:1998ia},  Drell Yan (DY) pair~\cite{Hammon:1996pw,Boer:1997bw,Boer:1999si} 
or isolated photon~\cite{Qiu:1991wg} production. 
For TMD observables,  one usually looks at processes immune to final-state radiations
which allows one to control the transverse momentum of the initial partons by measuring
momentum imbalances. In the case of \AFTER, the possible measurements go well 
beyond DY production and include pseudo-scalar quarkonium, 
quarkonium-pair and other associated production of colourless particles.

It is noteworthy that the TMD factorisation approach also allows one to investigate the structure of hadrons in a tridimensional momentum space 
(see \cite{Angeles-Martinez:2015sea} and references therein) in a rigorous and systematic 
way and it is not restricted to the study of STSAs. It provides theoretical tools to directly study 
not only the transverse-momentum distributions of the partons but also their
polarisation in both polarised and unpolarised nucleons. Such studies are in particular
related to azimuthal asymmetries in the final state.
At \AFTER~\cite{Brodsky:2012vg}, one can study them, for instance, via pseudo-scalar
 quarkonium production~\cite{Boer:2012bt,Signori:2016jwo}, associated quarkonium 
production~\cite{Dunnen:2014eta,Lansberg:2015lva} and, of course, DY pair production~\cite{Liu:2012vn,Anselmino:2015eoa,Kanazawa:2015fia}. 
We guide the reader to~\cite{Lansberg:2014myg,Massacrier:2015nsm} for more details.

With its high luminosity, a highly polarised target and an access towards the large momentum fraction $x^\uparrow$
in the target, \AFTER is probably the best set-up~\cite{Brodsky:2012vg,Lansberg:2016urh} to carry out an inclusive set of $A_N$ and azimuthal asymmetry measurements
both to improve existing analyses and to perform studies which are simply impossible elsewhere.
Let us recall that nearly nothing is known from the experimental side about the gluon Sivers effect 
(see e.g.~\cite{Boer:2015vso} for a recent review). The polarisation of not only hydrogen but also deuterium and helium targets 
allows for an even more ambitious spin program bearing on the neutron and spin 1 bound states~\cite{Boer:2016xqr}.

\section{Experimental implementation}

Two approaches for a fixed-target experiment at the LHC 
with a polarised target can be considered: one with a bent-crystal extracted
beam on a conventional polarised target, the other with an internal 
polarised gas target. Both have the virtue of being parasitic for they do not
affect the LHC performances. In the following, we will focus on the latter 
option\footnote{Such an internal gas-target option is in fact currently used by the LHCb 
collaboration but as a luminosity monitor~\cite{FerroLuzzi:2005em} (SMOG) 
initially designed to study the transverse size of the beam.  Despite its limited pressure
(about $10^{-7}$~mbar), it acts as an ideal demonstrator of such a solution over extended 
periods of time, without any interferences on the other LHC experiments. LHCb performed 
pilot runs with $p$ and Pb beams 
on a Ne gas target in 2012 and 2013. 
These pilot runs were followed by longer runs in 2015: $p$ on Ne (12 hours), He (8 hours) and Ar (3 days) 
as well as Pb on Ar (1 week) and $p$ on Ar (a few hours). 
As for now, the current gas pressure is limited by the pumping system.}
 which can directly be combined with existing detectors such as LHCb or ALICE. 
In particular, we will rely on the expected performances of a target system like the one of the 
HERMES experiment at DESY-HERA~\cite{Airapetian:2004yf},  
 as proposed in~\cite{Barschel:2015mka}.

\begin{table}[hbt!]
\begin{center}\small
\begin{tabular}{c c c c c c  c}
  Beam &  Target & $A$ & Areal density ($\theta$)   & $\cal{L}$ & $\int{\cal{L}}$    \\
       &         & &(cm$^{-2}$)    & ($\mu$b$^{-1}$.s$^{-1}$) & (fb$^{-1}$.y$^{-1})$  \\ \hline
  $p$     &   H      & 1 &$2.5\times 10^{14}$    & 900   &  9 \\
  $p$     &   D      & 2 &$3.2\times  10^{14}$ & 1200 & 12
\end{tabular}
\caption{Expected luminosities with an internal gas-target inspired by the HERMES experiment.}
\end{center}\vspace*{-0.5cm}
\end{table}

In such a case, the luminosity is given by the product
of the particle current $I$ and the areal density of the polarised storage cell target, $\theta$. 
The density results from many parameters, the flux of the polarised source 
injected into the target cell and the cell geometry are the most relevant ones~\cite{Steffens:2015kvp}. 
For a LHC compatible set-up~\cite{Steffens:2015kvp} and taking the HERMES-target-source flux, 
one can expect a density of  $2.5 \times 10^{14}$ cm$^{-2}$ for a 1 m long cell for H at 300 K
and, for D, $3.2 \times 10^{14}$~cm$^{-2}$. 
This is well below the densities which might affect the proton beam life time. 
The resulting luminosities with $I$ = 3.14 $\times 10^{18}$ $p^{+}$ s$^{-1}$ 
for the $p^{+}$ beam case are shown in Table 1. The limit of such a solution is 
essentially set by the number of minimum bias collisions by fill, that is
the number of proton ``consumed'' by the target as compared to the total number
 stored in the LHC.

In terms of the polarisation, by reusing a target like that of HERMES, 
the effective polarisation  of H or D~\cite{Barschel:2015mka} can be as high 
as 0.8. $^3$He can also be used. In the following, we will take an effective polariation of
$P=0.6$ as  a
working hypothesis which, from the number above, is definitely a 
conservative assumption. 

\section{Selected figures-of-merit for STSAs}\vspace*{-0.25cm}

In this section,  we quickly discuss figures-of-merit for open charm
and bottomonium production. Those for Drell-Yan and $J/\psi$ can be 
found in~\cite{Lansberg:2016urh}.  They both show that $A_N$ can
be measured close to the per cent level in regions where the 
expected effects can be as large as ten or even twenty per cent.

\subsection{STSAs in open (anti)-charm production}

Some years ago, it was argued~\cite{Kang:2008ih} that the study of STSAs 
in open heavy-flavour production in the RHIC energy domain  gives a direct access 
to the tri-gluon correlations functions appearing in the CT3 approach. 
Such correlations can be related to the gluon Sivers functions under some assumptions.
If STSAs for charm quarks and anti-quarks can separately be measured, 
this will also be a unique probe of $C$-parity odd twist-3 tri-gluon correlators~\cite{Ji:1992eu,Beppu:2010qn}.

\begin{figure}[!hbt]
\centering
{\includegraphics[width=0.45\textwidth]{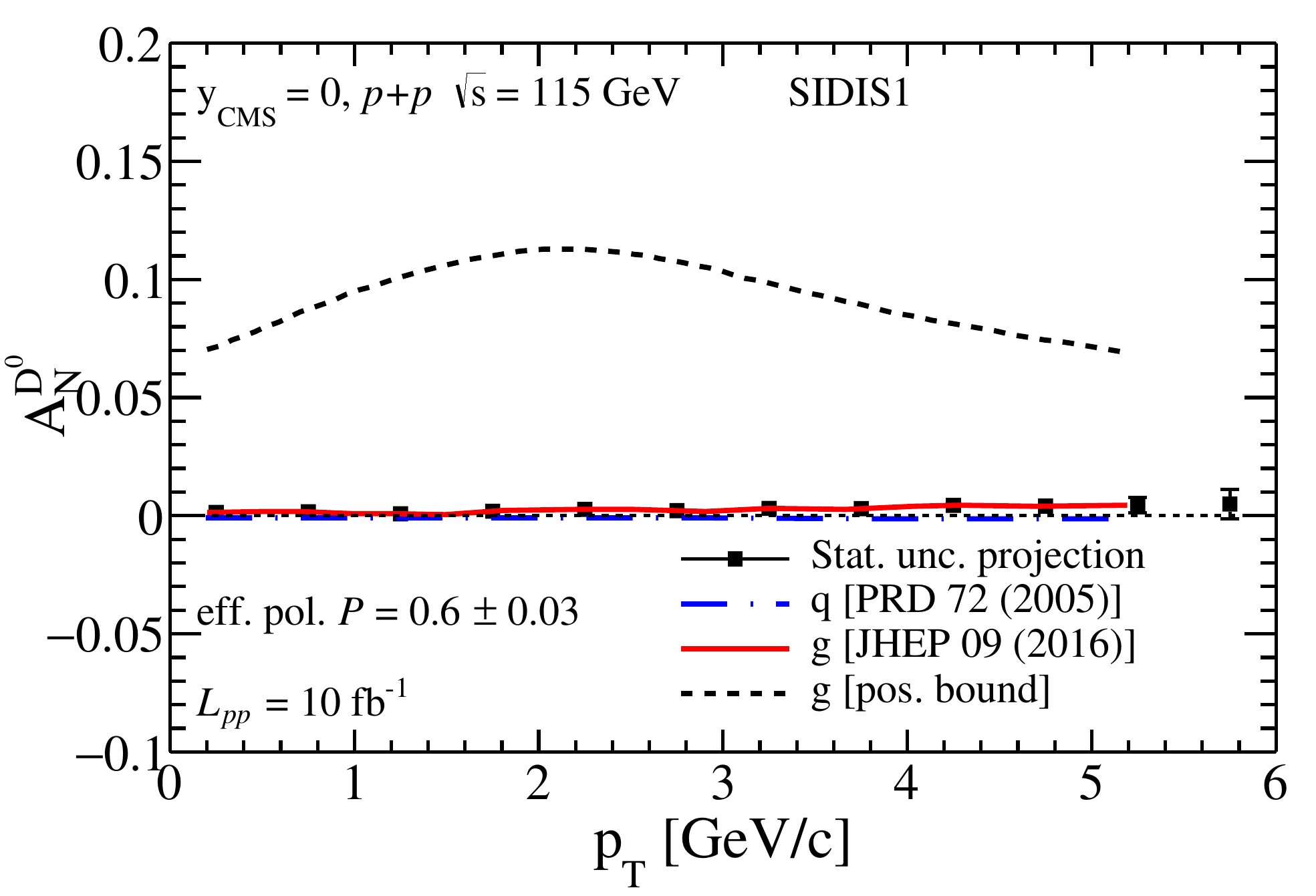}
}
{\includegraphics[width=0.45\textwidth]{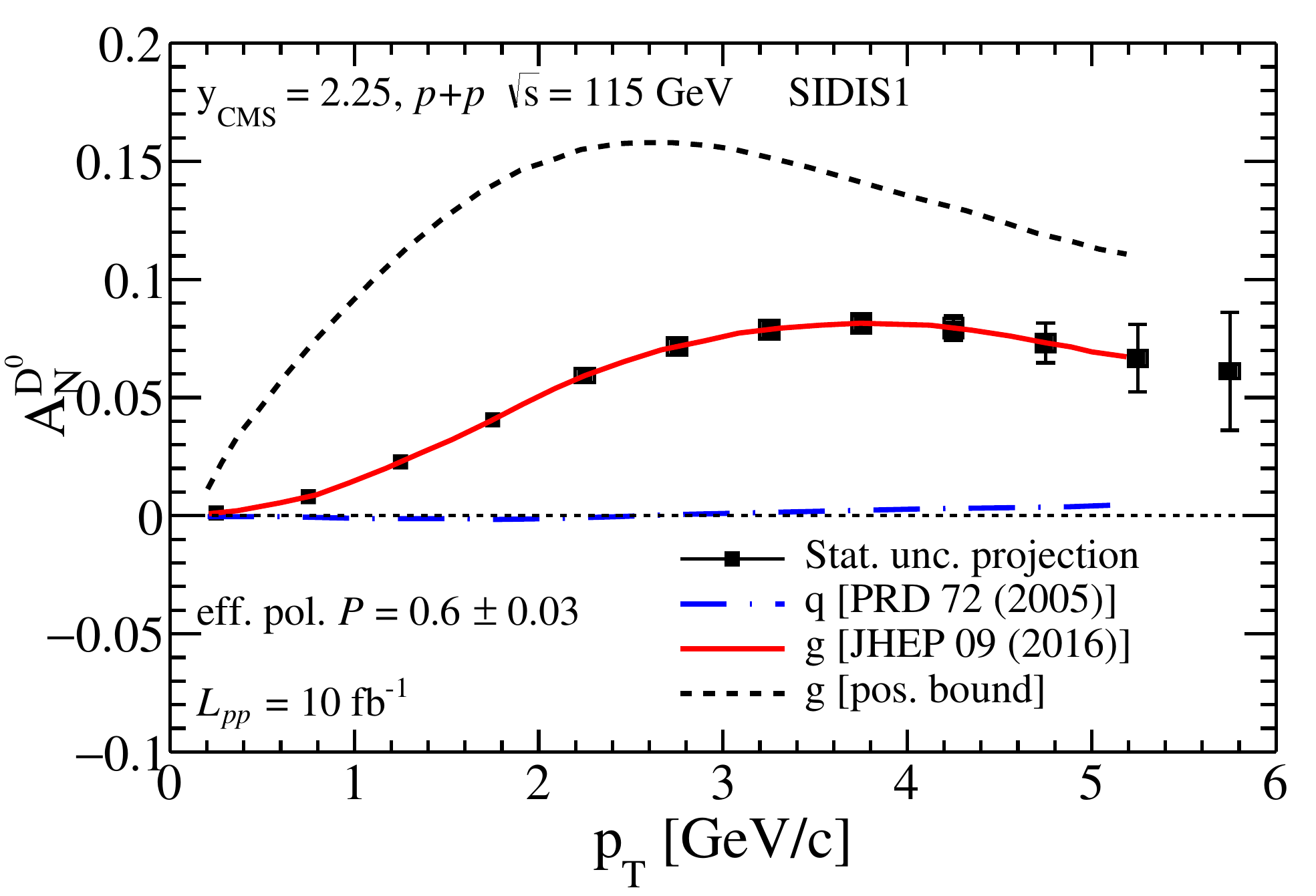}
}
\vspace*{-0.25cm}
\caption{\footnotesize{Projected $P_T$ dependence of $A_N$ for $D^0$ 
at $y_{\rm c.m.s.}=0$ ($y_{\rm c.m.s.}=2.25$) on the left (middle). 
}
}\label{fig:An-D0}
\end{figure}

\cf{fig:An-D0} (left \& middle) shows the projected statistical precision\footnote{In view of very low level
of background in various $D^0$ LHCb analyses, we neglected it here. We have also 
assumed LHCb-like acceptance and performances, see~\cite{Kikola:2015lka}.} for $D^0$ meson STSAs
at AFTER@LHC. It definitely lies at the per cent level. 
The expected magnitude is from~\cite{Anselmino:2004nk,D'Alesio:2015uta}.
As for now, such measurements are not planned elsewhere, certainly not
in the large $x^\uparrow$ region where the STSAs are expected to be the largest.

\subsection{STSAs in bottomonium production}

Bottomonia can also provide much information on the Sivers effects in the gluon 
sector, especially
via STSAs analyses of the $\Upsilon(1S)$, $\Upsilon(2S)$, $\Upsilon(3S)$ which
contain different amounts of $P$-wave feed-downs. We believe that
STSAs of $\Upsilon(2S)$ and $\Upsilon(3S)$ would not be accessible anywhere else than
at AFTER@LHC. Since only one particle is observed, these STSAs are preferably
treated in the CT3 approach that was, for instance, applied to the $\eta_c$ 
case~\cite{Schafer:2013wca} which could also be measured at AFTER@LHC. 
As for now, quantitive predictions for these STSAs are lacking even though
the qualitative conclusion of~\cite{Yuan:2008vn} should apply and the observation
of a non-zero STSA should already provide us with new information both 
on the quarkonium-production mechanisms~\cite{Andronic:2015wma,Brambilla:2010cs,Lansberg:2006dh}
and on the gluon Sivers effect~\cite{Boer:2015vso}.
For the figure-of-merit displayed on \cf{fig:An-Upsilon}, 
we have thus set the central value of the points to $A_N=0$. It clearly shows that, even for the
least produced $\Upsilon(3S)$, the statistical precision is better than 5\% taking
into account a full background simulation as discussed in~\cite{Massacrier:2015qba}.

\begin{figure}[!hbt]
\centering
{\includegraphics[width=0.5\textwidth]{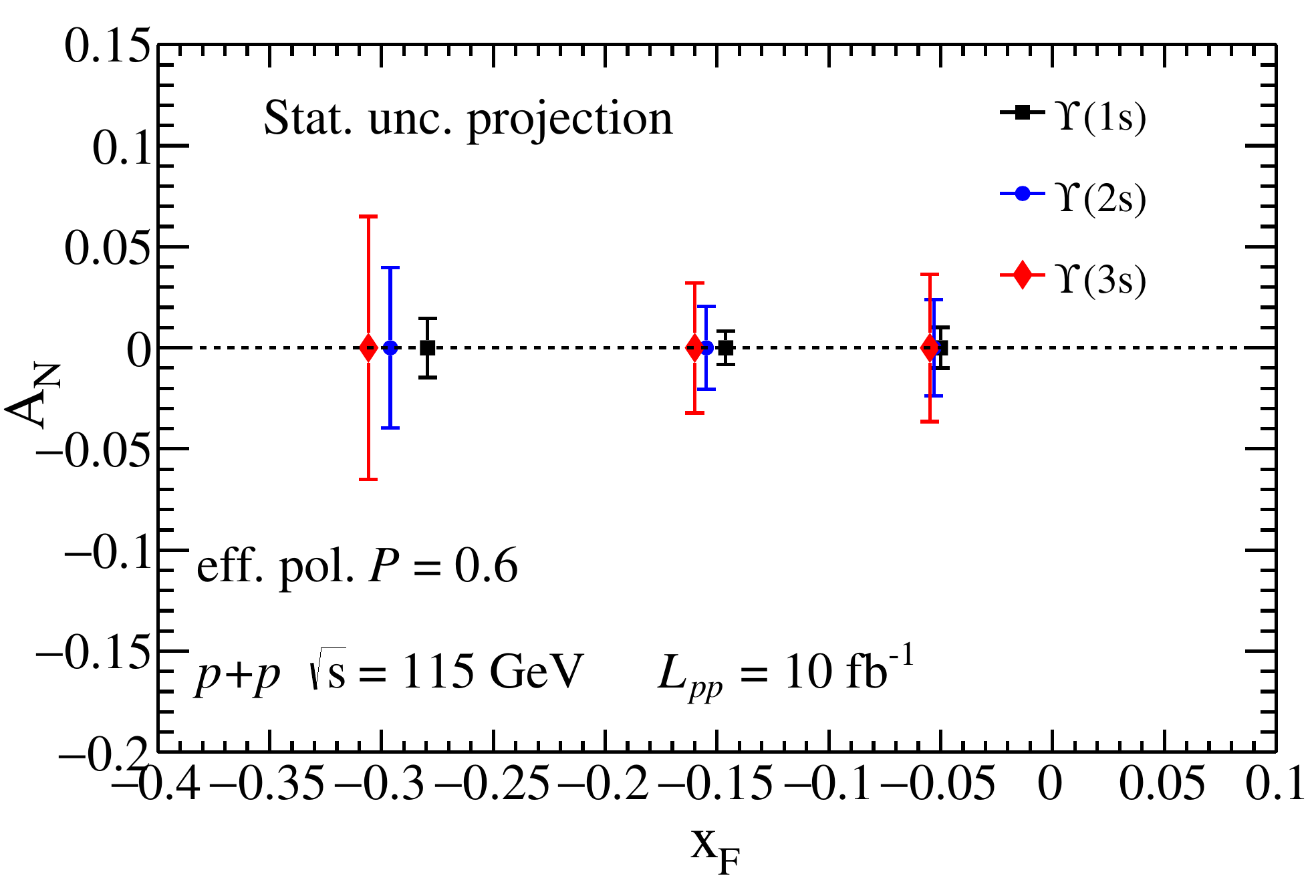}}
\vspace*{-0.25cm}
\caption{\footnotesize{Projected $x_F$ dependence of the statistical uncertainty on $A_N$ for 
$\Upsilon(1S)$, $\Upsilon(2S)$, $\Upsilon(3S)$ corresponding to 3
bins in rapidity ([2:3],[3:4],[4:5]).}}\vspace*{-0.25cm}
\label{fig:An-Upsilon}
\end{figure}

\section{Conclusion}

The combinaison of the TeV LHC proton beam, of a polarised internal gas-target system 
inspired by that of the HERMES experiment and a detector like LHCb (or ALICE)
opens the way to a number of outstanding STSA measurements at the per cent level,
most of which cannot be carried out at other facilities. 

As concluding remarks, we would like to emphasise two important facts: (i) at \AFTER, these STSAs 
would  naturally be measured at large $x^\uparrow$
where the (quark and gluon) Sivers effect is expected to be the largest and (ii)
such measurements have to be measured in $pp$ collisions
 as mandatory complementary pieces of information to similar studies in lepton-induced reactions
to perform quantitative tests of the generalised universality of the TMD-related observables, deeply connected to QCD.

\section*{Acknowledgements}
This research was supported in part by the French P2IO Excellence Laboratory, the French CNRS via the grants FCPPL-Quarkonium4AFTER \&
D\'efi Inphyniti--Th\'eorie LHC France and by the Department of Energy, contract DE--AC02--76SF00515.

\bibliographystyle{utphys}

\bibliography{AFTER-DIS2016-161016}

\end{document}